\documentclass[preprint, 12pt,nofootinbib]{revtex4-1}
\usepackage{graphicx}
\usepackage{cancel}
\usepackage{textcomp}
\usepackage{amsmath, amssymb, amsfonts, latexsym, epsfig}
\usepackage{mathrsfs}

\usepackage{bm}
\usepackage{times}
\usepackage{epsfig}
\usepackage{color}

\def\beq{\begin{equation}}
\def\eeq{\end{equation}}
\def\be{\begin{equation}}
\def\ee{\end{equation}}
\def\bea{\begin{eqnarray}}
\def\eea{\end{eqnarray}}
\def\to{\rightarrow}

\begin{document}
\title{Revisit to Non-decoupling MSSM}
\author{Jiwei Ke}
\author{Hui Luo}
\author{Ming-xing Luo}
\author{Kai Wang}
\author{Liucheng Wang}
\author{Guohuai Zhu}
\affiliation{ Zhejiang Institute of Modern Physics and Department of Physics, Zhejiang University, Hangzhou, Zhejiang 310027, CHINA}

\begin{abstract}
Dipole operator $\bar{s}\sigma_{\mu\nu}F^{\mu\nu}b$ requires the helicity flip in the involving quark states thus the breaking of chiral $U(3)_{Q}\times U(3)_{d}$. 
On the other hand, the $b$-quark mass generation is also a consequence of chiral $U(3)_{Q}\times U(3)_{d}$ symmetry breaking. Therefore, in many models, 
there might be strong correlation between the $b\to s\gamma$ and $b$ quark Yukawa coupling. 
In this paper, we use non-decoupling MSSM model to illustrate this feature.   
In the scenario,  the light Higgs boson may evade the direct search experiments at LEPII or Tevatron while the 125~GeV Higgs-like boson is identified as the heavy Higgs boson in the spectrum.  A light charged Higgs is close to the heavy Higgs boson which is of 125~GeV and
its contribution to $b\to s \gamma$ requires large supersymmetric correction with large PQ and $R$ symmetry breaking. 
The large supersymmetric contribution at the same time significantly modifies the $b$ quark Yukawa coupling.
 With combined flavor constraints $B\to X_{s}\gamma$ and $B_{s}\to \mu^{+}\mu^{-}$ and direct constraints on Higgs properties, we find best fit scenarios with light stop of $\cal O$(500~GeV), negative $A_{t}$ around -750~GeV and large $\mu$-term of 2-3~TeV. In addition, reduction in $b\bar{b}$ partial width may
 also result in large enhancement of $\tau\tau$ decay branching fraction. Large parameter region in the survival space under all bounds may be further constrained by $H\to \tau\tau$ if no excess of $\tau\tau$ is confirmed at LHC. We only identify a small parameter region with significant $H\to hh$ decay that is consistent with all bounds and reduced $\tau\tau$ decay branching fraction. In the end, if current dark matter mostly consists of neutralino, direct detection experiments like XENON100 also puts stringent bound over this scenario with light Higgs bosons. The light stops which are required by flavor constraints can further enhance the scattering cross section.
\end{abstract}

\maketitle

\section{Introduction}

A Higgs-like boson of 125~GeV has been discovered at the Large Hadron Collider (LHC) at CERN via two cleanest channels, the di-photon ($gg\to h\to \gamma\gamma$) and four-lepton ($gg\to h\to ZZ^{*}\to \ell^{+}_{i}\ell^{-}_{i}\ell^{+}_{j}\ell^{-}_{j}$ with $i,j=e^{\pm},\mu^{\pm}$) modes\cite{today}. Later both ATLAS and CMS collaboration also reported observations in the di-lepton  ($gg\to h\to WW^{*}\to \ell^{+}_{i}\nu_{i}\ell^{-}_{j}\bar{\nu}_{j}$) channels with the mass range consistent with the four-lepton measurement \cite{ww}. However, the confirmation of whether it is the Higgs boson of the standard model (SM) will require comprehensive and precise measurements of Higgs properties. The deviation of the Higgs couplings from the SM ones may imply the existence of the beyond SM physics, in particular, the excess in the di-photon channel
with $\sigma_{\rm obs.}/\sigma_{\rm SM} \sim 1.5-2.0$ at both ATLAS and CMS.
In all extension theories, additional charged and neutral scalars are inevitable. Therefore, searches of other Higgs-like states also provide direct test to models beyond SM physics. The LHC and Tevatron collaborations \cite{tevatron} have put stringent bounds over the SM Higgs, particularly heavy Higgs decaying into pure leptonic final states via $WW$ and $ZZ$. For instance, CMS collaboration has excluded the SM Higgs of 110-121.5~GeV and 128-600~GeV at 95\% C.L.  The LEPII experiments  also exclude the SM Higgs with mass lower than 114~GeV via $e^{+}e^{-}\to Zh$ channel. These bounds at the same time apply to various models with Higgs extension.

For two decades, weak scale supersymmetry has been the most elegant candidate to cancel the quadratic
divergence if the Higgs boson is indeed a fundamental scalar.
Within the supersymmetric framework, there exist several scenarios
where the di-photon decay branching fraction is enhanced,
for instance, models with light stau\cite{carlos} or light stop \cite{danhooper}.
Another particularly interesting region of non-decoupling limit  in minimal supersymmetric standard model (MSSM) has been discussed by various authors \cite{cp,Heinemeyer2011,Bottino,taohan,hagiwara,heinemeyer,Arbey,Belanger:2012tt,Drees}. It was observed that there might exist even lighter Higgs  $h$ which evades the search at LEP \cite{cp} due to suppressed $ZZh$ coupling and thus production of $Zh$. The light Higgs $h$ can then have $M_{h} < m_{Z}$ while the Higgs-like boson of 125~GeV can
be identified as the heavier degree of freedom $H$.
To reduce the $ZZh$ coupling $g_{ZZh}=\sin(\beta-\alpha)$ which is the vacuum expectation value ({\it vev})
of $h$, simple realization is to let $h$ be the $H_{d}$-like boson since
large $m_{t}$ naturally requires large $v_{u}$.  Given $h$ is a mixture state as $-\sin\alpha ({\rm Re}~H_{d}) + \cos\alpha ({\rm Re}~H_{u})$, this scenario prefers $\sin\alpha\simeq -1$ and large $\tan\beta$ which suppresses the $v_{d}$. In the limit of large $\tan\beta$ as $\sin\beta\to 1$, $\sin\alpha\to -1$ gives the $\sin(\beta-\alpha)$ approaches zero.
On the other hand, within MSSM, at tree level, the Higgs mass matrix gives
\beq
{\tan 2\alpha \over \tan 2\beta}= \frac{M^{2}_{A}+m^{2}_{Z}}{M^{2}_{A}-m^{2}_{Z}}
\eeq
Taking $M_{A}\to 0$ and the $\beta\to \pi/2$ as limit of large $\tan\beta$, one can get $\alpha\to -\pi/2$ which results in $\sin(\beta-\alpha)\to 0$ and reproduce the previous requirement. For $M_{A}\gtrsim 200$~GeV, the
$g_{ZZh}$ goes to the SM value. However, since the charged Higgs
state $H^\pm$ are at the similar scale as $M_{H}$ as $M^{2}_{H^{\pm}}= M^{2}_{A}+m^{2}_{W}$ at tree level, small $M_{A}$ leads to lighter $H^{\pm}$ which suffers from direct search bounds of light charged Higgs. Drell-Yan production of charged Higgs pair at LEP $e^{+}e^{-}\to H^{+}H^{-}$ put strict bounds as $M_{H^{\pm}}>80\sim 100$~GeV depends on its decay \cite{LEP2001}. Combining all the constraints, one expect an intermediate $M_{A}$ region around $m_{Z}$ scale to be consistent with the LEPII $Zbb$ search and charged Higgs search at LEP, Tevatron and LHC. In the limit of $M_{A}\to m_{Z}$, $h$ and $H$ masses at tree level are degenerate which is known as non-decoupling limit.

By requiring $M_{H}$ to be at 125~GeV, the first consequence of these non-decoupling scenarios is
that the charged Higgs is around similar scale. Charged scalar below top quark mass receives stringent bound from the ATLAS search of $t\to b H^{+}$ with $H^+ \to \tau^+ \nu$ requires the BR$(t\to b H^{+})\times \text{BR}(H^{+}\to \tau^{+}\nu_{\tau})<1\sim5\%$ for mass range $M_{H^{\pm}}$ in between 90 and 160~GeV\cite{Aad:2012tj}.
In the conventional Two-Higgs-Doublet models (2HDM) such a
light charged Higgs suffer severe constraints due to flavor violation processes\cite{flavor}. For example, one might be concerned by $B_{u}\to \tau \nu_{\tau}$ and $B\to D^{(*)}\tau\nu_{\tau}$ decays which receive charged Higgs contributions at the tree-level.
The two most sensitive parameters involved in Higgs interaction are $M_{A}$ and $\tan\beta$.
As we argued, $M_{A}$ is taken to be not much heavier than $m_{Z}$ and LEP2 $Zh$ search prefers a relatively
large $\tan\beta$. In addition, as we will show later, the recent search of $t\to b H^{+}$ at the LHC restricts
$\tan\beta\sim 10$ in non-decoupling region. For $B_{u}\to \tau\nu_{\tau}$ decay, The $W^{\pm}$-mediated SM contribution is helicity suppressed.
Therefore, even though the charged scalar is somewhat heavier, its contribution could be comparable to the SM part if $\tan\beta$ is not small \cite{Hou,Akeroyd,Isidori}:
\beq
\frac{\text{BR}(B^+ \to \tau^+ \nu)_{\rm MSSM}}{\text{BR}(B^+ \to \tau^+ \nu)_{\rm SM}}\simeq \left( 1- \frac{m_B^2}{M_{H^+}^2}\tan^2\beta\right)^2
\eeq
where the MSSM corrections to the down quark and lepton mass matrix have been neglected, which is safe for $\tan\beta\sim 10$.
For $M_{H^+}$ lies around $120 \sim 150$ GeV, the MSSM prediction would be about $20\% \sim 30\% $ smaller than the SM result of   $(0.95\pm0.27)\times 10^{-4}$. While the experimental world average is $(1.65\pm0.34)\times 10^{-4}$ before 2012 \cite{HFAG}, Belle updated their measurement at ICHEP2012 with much smaller value $0.72^{+0.29}_{-0.27} \times 10^{-4}$ for hadronic tag of $\tau$ \cite{Nakao}. So in the non-decoupling limit, a light
charged Higgs with $\tan\beta \sim 10$ is well consistent with the new Belle measurement. Similarly, the charged Higgs contribution to
$B\to D^{(*)}\tau\nu_{\tau}$ decays are not very significant in the interesting region of $M_{H^+}$ and $\tan\beta$. Therefore we will not discuss the bounds from $B^+ \to \tau^+\nu$ and $B\to D^{(*)}\tau\nu_{\tau}$ decays further in our study.

On the other hand, the penguin $b\to s$ processes are also sensitive to the charged Higgs effects. Generally, $b\to s \gamma$
and $B_{s}\to \mu^{+} \mu^{-}$ are two most stringent constraints. But choosing appropriate MSSM parameters, supersymmetric contributions may cancel part of the SM and charged Higgs amplitudes \cite{yangjinmin}. For example, $b\to s$ transition mediated by the scalar top quark (stop) loop in MSSM may cancel the top quark loop in SM and 2HDM in some parameter region. One can thus expect that light stop in MSSM may significantly reduce the flavor violation \cite{qaisar}.
In this paper, we start with this argument and study whether scenarios with light stop can resolve the
tension in flavor physics due to the light charged Higgs $H^{\pm}$.

Search of $\mu\to e\gamma$ at the MEG experiment will soon
reach BR$(\mu\to e\gamma)\simeq 1\times 10^{-13}$.
The one loop contribution from charged state to $\mu\to e\gamma$
is suppressed by small lepton masses and additional helicity-flip.
The largest contribution in
Higgs mediated $\mu\to e \gamma$ is usually the Barr-Zee two-loop
effects involving the charged scalar coupling to a top-bottom loop.
However, \cite{hisano} shown that the charged Higgs contribution
only reach the sensitivity for $\tan\beta$ of 60 for $M_{A}$ of 100~GeV where
the $\tan\beta$ is much larger than what is considered in non-decoupling scenarios.

With conserved R-parity, the thermal relic abundance of the lightest neutralino (LSP) can often be identified with dark matter (DM), consistent with the current cosmological observations. In recent years, direct detection of weakly interacting (WIMP) DM particle through the DM scattering with nuclei has excluded large parameter space of supersymmetric DM and put stringent bound on many models. The latest bound from XENON100 is about $5\times 10^{-9}$ pb for DM mass around $200$ GeV \cite{Xenon100}. Neutral Higgs states $h,H$ can also mediate the scattering between DM and nuclei which is of $1/M^{4}_{h,H}$.  Then the second consequence of non-decoupling scenarios is that the spin-independent scattering is significantly enhanced by the interaction through neutral Higgs $H,A$ of $\cal O$(100~GeV) \cite{directdetection}.
Therefore, models with only neutralino DM in the non-decoupling MSSM suffer stringent constraints from direct detection experiments. In addition, light stop which may significantly improve the flavor physics behavior of non-decoupling MSSM as argued above, would further enhance the scattering of DM and nuclei and put stronger bound on non-decoupling scenarios with only neutralino DM \footnote{If the DM is not dominated by the neutralino component, the bound can be evaded.}.

In the next section, we discuss some general constraints on the non-decoupling scenarios
and the scan results. Then we discuss in details the physics interpretation of the scan results,
in particular, light stop contribution to cancel light charged Higgs and its implication to $M_{H}$,
di-photon, di-tau decay and the direct detection experiments of neutralino dark matter. 
We then conclude in the final section.

\section{General Constraints and their implications}

In this section, we first scan the parameter space with focus on non-decoupling region with $M_{A}$ is at ${\cal O}(m_{Z})$ then discuss in details the physics interpretation of scan results.

Latest data from the LHC require the resonance to be at 125~GeV with di-photon decay enhanced with respect to the SM prediction. We therefore impose the selection rules as
\begin{itemize}
\item $M_{H}: 125\pm 2$~GeV;
\item $R_{\gamma\gamma}=\sigma^{\gamma\gamma}_{\rm obs}/\sigma^{\gamma\gamma}_{\rm SM}: 1\sim 2$;
\item Combined direct search bounds from HiggsBound3.8.0;
\item BR$(B\to X_{s}\gamma)<5.5\times 10^{-4}$;
\item BR$(B_{s}\to \mu^{+}\mu^{-})<6\times 10^{-9}$~.
\end{itemize}

Without loss of generality, we fix masses of the following sfermions as
\beq
M_{\tilde{Q}_{1,2}}=M_{\tilde{u}_{1,2}}=M_{\tilde{d}_{1,2,3}}=M_{\tilde{L}_{1,2,3}}=M_{\tilde{e}_{1,2,3}}=1~\text{TeV}~,
\eeq
and the gauginos as
\beq
M_{1}=200~\text{GeV}, M_{2}=400~\text{GeV}, M_{3}=1200~\text{GeV}~.
\eeq
As argued, our study focus on the flavor constraints of the non-decoupling MSSM and $b\to s$ transitions like
$B\to X_{s}\gamma$ and $B_{s}\to \mu^{+}\mu^{-}$ provide the most severe constraints.
Light stop usually helps to cancel the charged Higgs contribution in $b\to s$ transition. On the other hand, for light stop below 500~GeV,  we find that the gluon fusion production of $H$ is  suppressed significantly with respect to the SM value due to the cancellation between top squark and top quark in the loop. Thus, for light stop ($M_{\tilde{t}}<500$~GeV), it is difficult to achieve enhanced di-photon.  
For comparison, we take the third generation up quark masses as
\beq
M_{\tilde{Q}_{3}}=M_{\tilde{t}}=500~\text{GeV}~~~~\text{and a second group with 1~TeV}~.
\eeq

We do the scan over four parameters \footnote{We confine ourselves to $M_{A}\lesssim 150$~GeV for larger splitting between $h$ and $H$ which can reduce the $\tau\tau$ decay branching ratio. Details is discussed later.  }
\bea
M_{A} &:& 95\sim 150~\text{GeV}\nonumber\\
\tan\beta&:& 1\sim30\nonumber\\
\mu&:& 200~\text{GeV}\sim 3~\text{TeV}\nonumber\\
A_{u}=A_{d}=A_{\ell}&:&-3\sim 3~\text{TeV}~.
\eea
Discussed by many authors\cite{carlos}, light stau states may significantly enhance the di-photon rate of the Higgs-like boson
decay which are observed by both ATLAS and CMS collaborations. On the other hand, we don't require much stronger
di-photon bound as $R_{\gamma\gamma}$ to be the experimental preferred central value of 1.5. Stau states are irrelevant
to the flavor constraints from $b\to s$ transition but only give minor change to the Higgs boson mass. Therefore, we don't take light stau in the study.

\begin{figure}[h]
\includegraphics[scale=1,width=7cm]{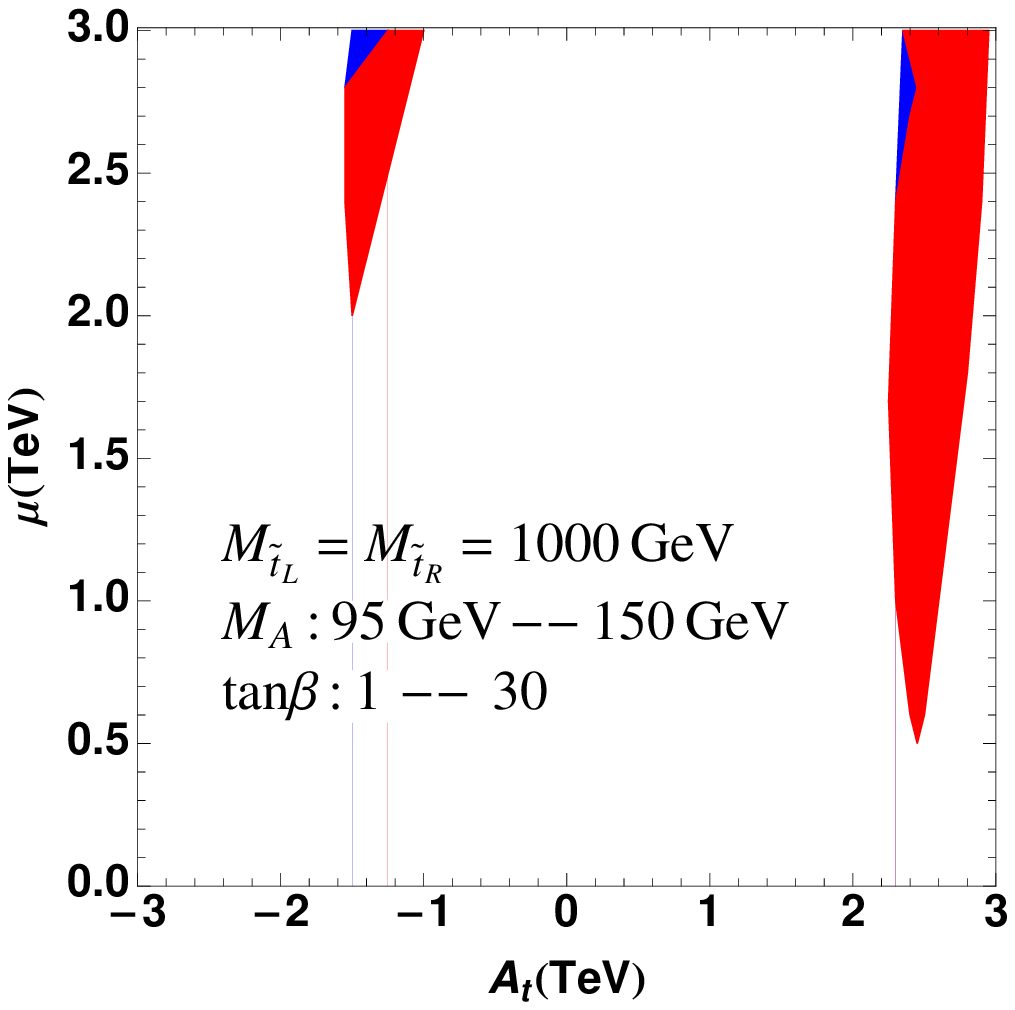}
\includegraphics[scale=1,width=7cm]{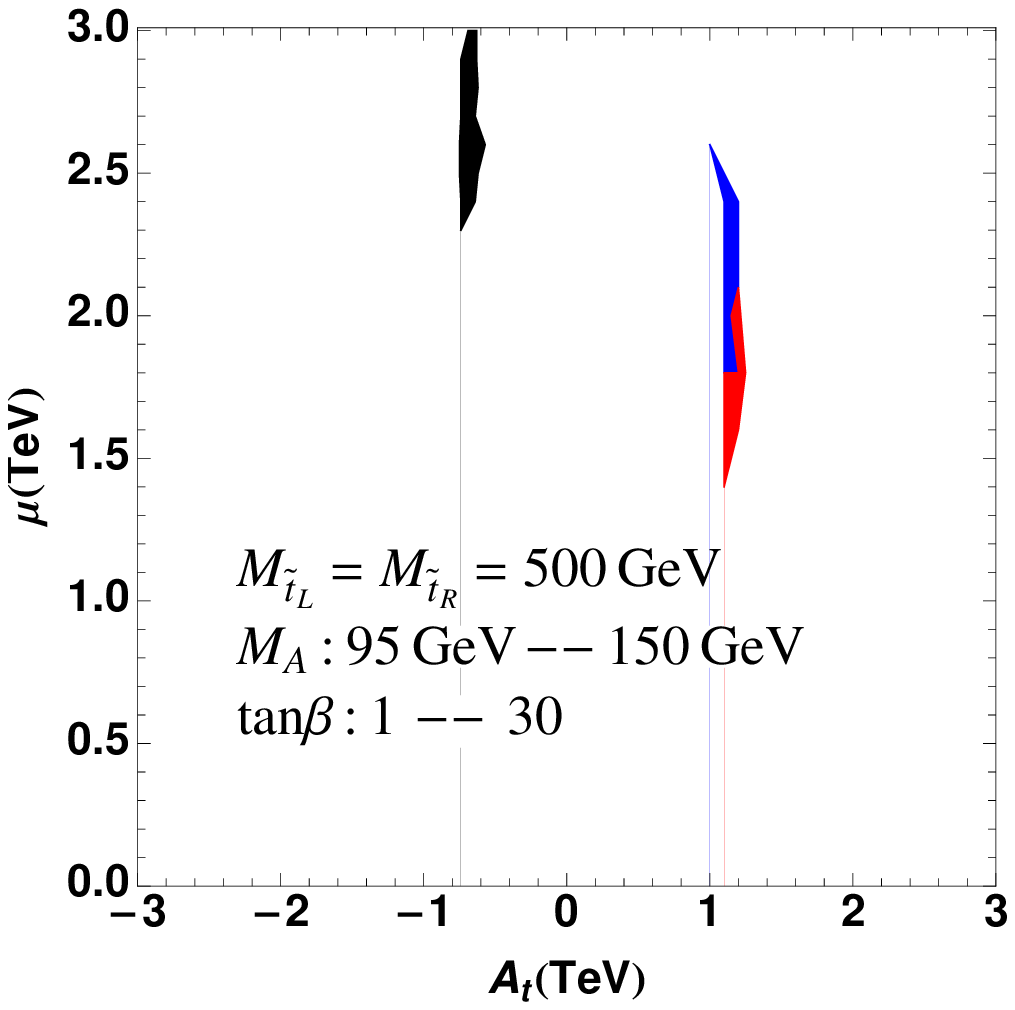}
\caption{Scan Results in [$A_t$, $\mu$] plane. The heavy (light) stop scenario with $M_{\tilde{Q}_{3}}=M_{\tilde{t}}=1~(0.5)$~TeV is shown in the left (right) plot. The red region pass the direct search bounds from HiggsBounds with
a heavy CP-even Higgs $M_H=125 \pm 2$ GeV and an enhanced diphoton rate $1<R_{\gamma \gamma}<2$. The blue region pass in addition the constraint of BR$(B\to X_s \gamma)$, while the black region pass all the constraints, including further the restriction of BR$(B_s \to \mu^+ \mu^-)$.  }
\label{scan}
\end{figure}

We use {\it FeynHiggs 2.9.2}~\cite{feynhiggs} \footnote{In this scan, we take the pole mass of $m_{t}$ instead of the running $m_{t}$ mass. The survival parameter region after scan may be shifted by a few percent. } with {\it HiggsBounds 3.8.0}~\cite{higgsbounds} and {\it SUSY\_Flavor 2.01}~\cite{Crivellin:2012jv} to perform the scan here. 
Figure \ref{scan} shows the scan results in 2D-plot of $A_t$ and $\mu$. $M_{A}$ and $\tan\beta$ are also varied but
aren't shown in the figures. Figure \ref{scan}-(a) is the heavy stop scenario with $M_{\tilde{Q}_{3}}=M_{\tilde{t}}=1$~TeV
and (b) is the light stop scenario with $M_{\tilde{Q}_{3}}=M_{\tilde{t}}=500$~GeV. Points in red region pass the direct search bounds from HiggsBounds with a heavy CP-even Higgs $M_H=125 \pm 2$ GeV and an enhanced diphoton rate $1<R_{\gamma \gamma}<2$.    
The points in blue region pass in addition the constraint of BR$(B\to X_s \gamma)$, while the points in black region pass all the constraints, including further the restriction of BR$(B_s \to \mu^+ \mu^-)$.  

The scenario with heavy stop can survive the $B\to X_{s}\gamma$ constraints. However, none of the
scanned points can pass the $B_{s}\to \mu^{+}\mu^{-}$. 
In the case of light stop of 500~GeV, we find a small survival parameter region
with negative $A_{t}$ around 750~GeV and large $\mu$-term between 2 to 3~TeV.
In the following subsections, we discuss in details the physics implications of the scanned results.

\subsection{$b\to s\gamma$ and $B_{s}\to \mu^{+}\mu^{-}$}

$b\to s\gamma$ and $B_{s}\to \mu^{+}\mu^{-}$ turns out to be the most stringent flavor physics
bounds in the non-decoupling limit.
The helicity for the involved quark states must be flipped in $b\to s\gamma$.
Hence, both chiral symmetry $U(3)_{Q}\times U(3)_{d}$ and
electroweak symmetry $SU(2)_{L}\times U(1)_{Y}$ must be broken.
The SM  $b\to s$ transition is mediated by the charged weak boson $W^{-}$ and
only left-handed quarks are involved in the weak interaction. Consequently,
$b\to s\gamma$ is suppressed by mass insertion of bottom quark mass $m_{b}$ in the SM.
In MSSM, the charged Higgs $H^{-}$-top quark loop contribution to $b\to s\gamma$ is also
suppressed by $m_{b}$ insertion, and has the same sign as the SM amplitude. Besides the above contributions, squarks
can also generate $b\to s\gamma$ in MSSM which may not flip the helicity of
the involved quark states, for instance, loops with right-handed stop-Higgsino
($\tilde{t}_{R}-\tilde{H}_{u}$) or left-handed stop-Wino ($\tilde{t}_{L}-\tilde{W}$). Therefore the squark contributions, in particular the top squark ones, are not necessarily suppressed by $m_b$, which is helpful to cancel the SM and charged Higgs amplitudes with appropriate MSSM parameters. Consequently, scalar top quark with small $M_{\tilde{t}}$, say $\sim 500$ GeV, could significantly reduce $b\to s$ transition.

The squark contributions can be decomposed into chargino penguins, wino penguins and gluino penguins. 
Chargino penguins contain $\tan\beta$-enhanced term which arises from $v_{u}$ insertion in $Q d^{c} \langle H^{*}_{u}\rangle$. The term explicitly breaks Peccei-Quinn symmetry as well as $R$-symmetry and is proportional to $\mu A_{t}$. This contribution would destructively interfere with the SM and charged Higgs amplitudes in case of $\mu A_t<0$ \cite{Barbieri,Carena:1994bv}. In our study, gluino penguins are also important as they contain terms enhanced by $\mu \tan\beta$ and terms chirally enhanced by $m_{\tilde g}/m_b$. Numerically, we use the {\it FeynHiggs} program to get the Non-MFV result of BR$(B \to X_s \gamma)$. The experimental world average of this process is $(3.43 \pm 0.22)\times 10^{-4}$ \cite{HFAG}, while the SM prediction up to NNLO perturbative QCD corrections is $(3.15\pm 0.23)\times 10^{-4}$ \cite{Misiak:2006zs}. However, $B \to X_s \gamma$ decay is evaluated only at NLO in the FeynHiggs program, which produces the SM result as $3.8 \times 10^{-4}$. This is about $30\%$ larger than the NNLO SM prediction. Taking this and the theoretical and experimental uncertainties into account, we require loosely BR$(B \to X_s \gamma)_{MSSM}<5.5\times 10^{-4}$ as the selection rule in the scan.

In the SM, BR($B_s \to \mu^+ \mu^-$) is strongly helicity suppressed by the small muon mass as $m_\mu^2/m_{B_s}^2$, which leads to a tiny branching ratio of $(3.27 \pm 0.23)\times 10^{-9}$ \cite{Buras:2012ru}. However, it is well known that the MSSM contributions to this decay could be enhanced several orders of magnitude larger than the SM prediction in large $\tan\beta$ limit, as the leading contribution of Higgs penguin diagrams to the branching ratio are proportional to $\tan^6 \beta$. In our study, $\tan\beta \sim 10$ is not very large, so all the 1-loop diagrams have to be considered, including the charged Higgs diagrams which is enhanced up to $\tan^2 \beta$ at the amplitude level. Notice that $B_s \to \mu^+ \mu^-$ decay is even more sensitive to the MSSM parameters in the non-decoupling limit as the neutral Higgs bosons are all light. Experimentally, a combined search of ATLAS, CMS and LHCb has set the upper limit of $4.2 \times 10^{-9}$ \cite{CMS-PAS} for time integrated branching ratio. As pointed out in \cite{DeBruyn:2012wj,DeBruyn:2012wk}, this upper limit should be reduced by about $10\%$ when compared with the theoretical calculation. Numerically, we use the SUSY\_FLAVOR program \cite{Crivellin:2012jv} to get the complete NLO result of BR$(B_s \to \mu^+ \mu^-)$. However, we notice that SUSY\_FLAVOR evaluates this branching ratio to be $4.8 \times 10^{-9}$ in the SM. This is about $50\%$ larger than the SM prediction of $(3.27 \pm 0.23)\times 10^{-9}$ in \cite{Buras:2012ru}, probably mainly due to different choice of hadronic parameters. Taking this into account, we set the corresponding selection rule to be BR$(B_s \to \mu^+ \mu^-)_{MSSM}<6\times 10^{-9}$ in the scan.

In Fig. \ref{scan}, the black region which satisfy all the constraints give $10^4\text{BR}(B \to X_s \gamma)_{MSSM}$ in the region [$4.9$, $5.3$] and $10^9\text{BR}(B_s \to \mu^+ \mu^-)_{MSSM}$ in the region [$2.3$, $4.3$]. Notice that BR$(B \to X_s \gamma)$ is always larger than the SM prediction, which is mainly due to the enhancement of light charged Higgs. For BR$(B_s \to \mu^+ \mu^-)$, it is always somewhat smaller than the SM prediction.

\subsection{Higgs mass and its decay properties}

\begin{figure}[h]
\includegraphics[scale=1,width=7cm]{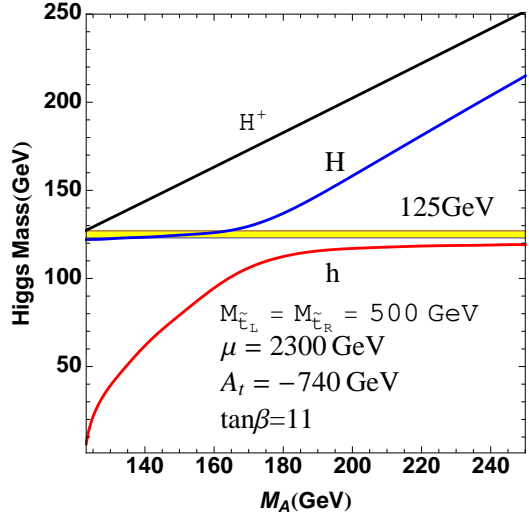}
\caption{$M_{h,H,H^{\pm}}$ vary with respect to $M_{A}$ for $M_{\tilde{t}}=500$~GeV, $A_{t}=-740$~GeV, $\tan\beta=11$, $\mu=2300$~GeV. }
\label{mh}
\end{figure}

We discuss the mass spectrum of the Higgs bosons in non-decoupling
MSSM and its decay properties in this section. More general discussion can be found in \cite{physicsreport}.
In particular, we focus on the parameter region that minimizes the flavor violation
in $b\to s$ transition. Combined constraints from $B\to X_{s}\gamma$ and $B_{s}\to \mu^{+}\mu^{-}$,
we take light stop of $M_{\tilde{t}}\sim 500$~GeV with negative $A_{t}$ of ${\cal O}(-750~\text{GeV})$ and large $\mu$-term of 2-3~TeV.
MSSM contains two $SU(2)_{L}$ doublets $H_{u}$ and $H_{d}$ with the ratio of their {\it vev}s $\tan\beta= v_{u}/v_{d}$.
To evade LEPII bounds, non-decoupling limit corresponds to a region of much lighter $H_{d}$ state with small {\it vev} in
the spectrum.
After spontaneously electroweak symmetry breaking, MSSM gives rise to five physical states
of Higgs bosons, the two CP even scalar $h,H$ with one CP odd scalar state $A$ and charged scalars
$H^{\pm}$. The two CP even scalar bosons $h,H$ arise from mixing of the real gauge eigenstates $(\text{Re}~H_{d},\text{Re}~H_{u})$,
\begin{equation}
\left(\begin{array}{c}
h\\H
\end{array}\right)=
\left(\begin{array}{cc}
-\sin\alpha &\cos\alpha\\
\cos\alpha & \sin\alpha
\end{array}\right)
\left(\begin{array}{c}
\text{Re}~H_{d}\\
\text{Re}~H_{u}\end{array}\right)~.
\end{equation}
After diagonalizing the general mass matrix of neutral Higgs
\begin{equation}
\mathcal{M}^{2}=\left(\begin{array}{cc}
\mathcal{M}_{11}^{2} & \mathcal{M}_{12}^{2}\\
\mathcal{M}_{21}^{2} & \mathcal{M}_{22}^{2}
\end{array}\right),
\end{equation}
the masses of two CP-even Higgs are
\begin{equation}
\begin{cases}
M_{h}^{2} & =\mathcal{M}_{11}^{2}\sin^{2}\alpha+\mathcal{M}_{22}^{2}\cos^{2}\alpha-\mathcal{M}_{12}^{2}\sin2\alpha,\\
M_{H}^{2} & =\mathcal{M}_{11}^{2}\cos^{2}\alpha+\mathcal{M}_{22}^{2}\sin^{2}\alpha+\mathcal{M}_{12}^{2}\sin2\alpha,
\end{cases}
\label{matrix}
\end{equation}
To illustrate the feature, we take the limit of $\sin(\beta-\alpha)\to 0$ which is the vanishing limit of $g_{ZZh}$ to completely suppress the $Zh$ production at  LEPII. As a result of $\sin\alpha\to -1$
and $\sin\beta\to 1$, we have
\beq
\begin{cases}
M_{h} & \simeq {\cal M}_{11}\\
M_{H} & \simeq {\cal M}_{22}
\end{cases}
\eeq
Radiative corrections to the elements in mass matrix Eq. \ref{matrix} are given in \cite{Carena:1995bx}. We list
the most relevant ${\cal M}_{22}$ in Eq. \ref{m22}
\bea
\label{m22}
M^{2}_{H}\simeq {\cal M}^2_{22} & \simeq & M_A^2 \cos^2\beta + m_Z^2 \sin^2\beta
\left(1 - \frac{3}{8 \pi^2} y_t^2 t \right)
\nonumber\\
& + &\frac{y_t^4 v^2}{16 \pi^2} 12 \sin^2\beta \left\{
t \left[ 1 + \frac{t}{16 \pi^2} \left( 1.5 y_t^2 + 0.5 y_b^2 -
8 g_3^2 \right)
\right]
\right.
\nonumber\\
& + & \left.
\frac{{A}_t\tilde{a}}{M^{2}_{SUSY}}\left(1  - {{A}_t\tilde{a} \over 12 M^{2}_{SUSY}}\right)
\left[ 1 + \frac{t}{16 \pi^2} \left( 3 y_t^2 +  y_b^2 - 16 g_3^2 \right)
\right] \right\}
\nonumber\\
& - &   \frac{v^2 y_b^4}{16 \pi^2} \sin^2\beta \frac{{\mu}^4}{M^{4}_{SUSY}}
\left[ 1 + \frac{t}{16 \pi^2} \left( 9 y_b^2 - 5 y_t^2 - 16 g_3^2 \right)
\right]
+ {\cal O}(y_t^2 m_Z^2)
\label{mhh}
\eea
where $g_3$ is the QCD running coupling constant, $y_t$
and $y_b$ are the top and bottom Yukawa couplings.  $M_{SUSY}$ is
the arithmetic mean of top squark masses $M_{\tilde{t}}$. $A_{t}$
is the SUSY breaking $A$-term associated with top squark and
$\mu$ is the Higgsino mass parameter. $t$ is defined as $\ln(M_{SUSY}^2/m_t^2)$ and
\beq
\tilde{a}\equiv A_t - \mu/\tan\beta~.
\eeq
In Eq.\ref{mhh}, only the leading terms in powers of $y_b$ and $\tan\beta$ have been retained.
Even though the Eq.\ref{mhh} is only valid in the limit of small splittings
between the running stop masses, it shows the qualitative feature for
how couplings to stop and sbottom modify the Higgs masses.
$M_{A}$ is $\cos\beta$ dependent which is suppressed in the limit of large $\tan\beta$.
Therefore, the $M_{H}$ is not very sensitive to $M_{A}$ and with $\tan\beta\simeq 10$,
varying $M_{A}$ by 100~GeV results in 10~GeV difference in $M_{H}$.
Unlike the $m^{\rm max}_{h}$ scenario with $\tilde{a}=\sqrt{6}M_{SUSY}$ which is usually used in many studies,
minimization of the flavor violating $b\to s$ transition leads to our best fit parameter region around
\beq
A_{t}\sim -750~\text{GeV}, M_{\tilde{t}}\sim 500~\text{GeV}, \mu\sim 2000-3000~\text{GeV}~.
\eeq
The particular choices of $A_{t}$ and $M_{\tilde{t}}$ significantly modifies the Higgs boson masses through radiative corrections.
In our studies, we use the {\it FeynHiggs} program to compute the mass spectrum of Higgs in which
full radiative corrections of Higgs masses have been implemented \cite{feynhiggs}.
Figure \ref{mh} show how the $M_{h,H,H^{\pm}}$ vary with respect to $M_{A}$ for one of our benchmark points $\tan\beta=11$,  $M_{\tilde{t}}=500$~GeV, $A_{t}=-740$~GeV and $\mu=2300$~GeV.
For a large range of $M_{A}$, $M_{H}$ is around 125~GeV. Non-decoupling limit of nearly
degenerate $h,H$ lies near $M_{A}\sim 160$~GeV.

Since $H$ is mostly $H_{u}$ with large $\tan\beta$, the $v_{u}$ dominates the
electroweak symmetry breaking $v$. The couplings between $H$ and $W^{+}W^{-}$
and top quark $t$ are similar to their SM values. Since the di-photon decay is dominated
by the $W$-boson contribution, the di-photon decay partial width
is not changed significantly from the SM $\Gamma_{\rm SM}(H\to \gamma\gamma)$.
However, di-photon decay branching fraction BR($H\to \gamma\gamma$) may still
be enhanced due to decrease of $H$ total width. At 125~GeV, $H\to b\bar{b}$
and $H\to WW^{*}$ $H\to ZZ^{*}$ dominate the $H$ decay. Since $H$ is mostly $H_{u}$-like,
$H$ coupling to $b$ is naturally suppressed. 
Given $v_{u}\sim v$, $g_{HZZ}$ and $g_{HWW}$ are not significantly
changed from the SM $g^{\rm SM}_{hZZ}$ and $g^{\rm SM}_{hWW}$. The partial
widths of  $H\to WW^{*}$ and $H\to ZZ^{*}$ are indifferent from the SM values.
With the reduction in $H\to b\bar{b}$, the increase of $H\to WW^{*}$ and $H\to ZZ^{*}$
are inevitable. Therefore, light stau states in the spectrum can improve the
di-photon behavior $R_{\gamma\gamma}$ and reduce the tension in
increasing ${ZZ^{*}}$ or ${WW^{*}}$.

\begin{figure}[h]
\includegraphics[scale=1,width=7cm]{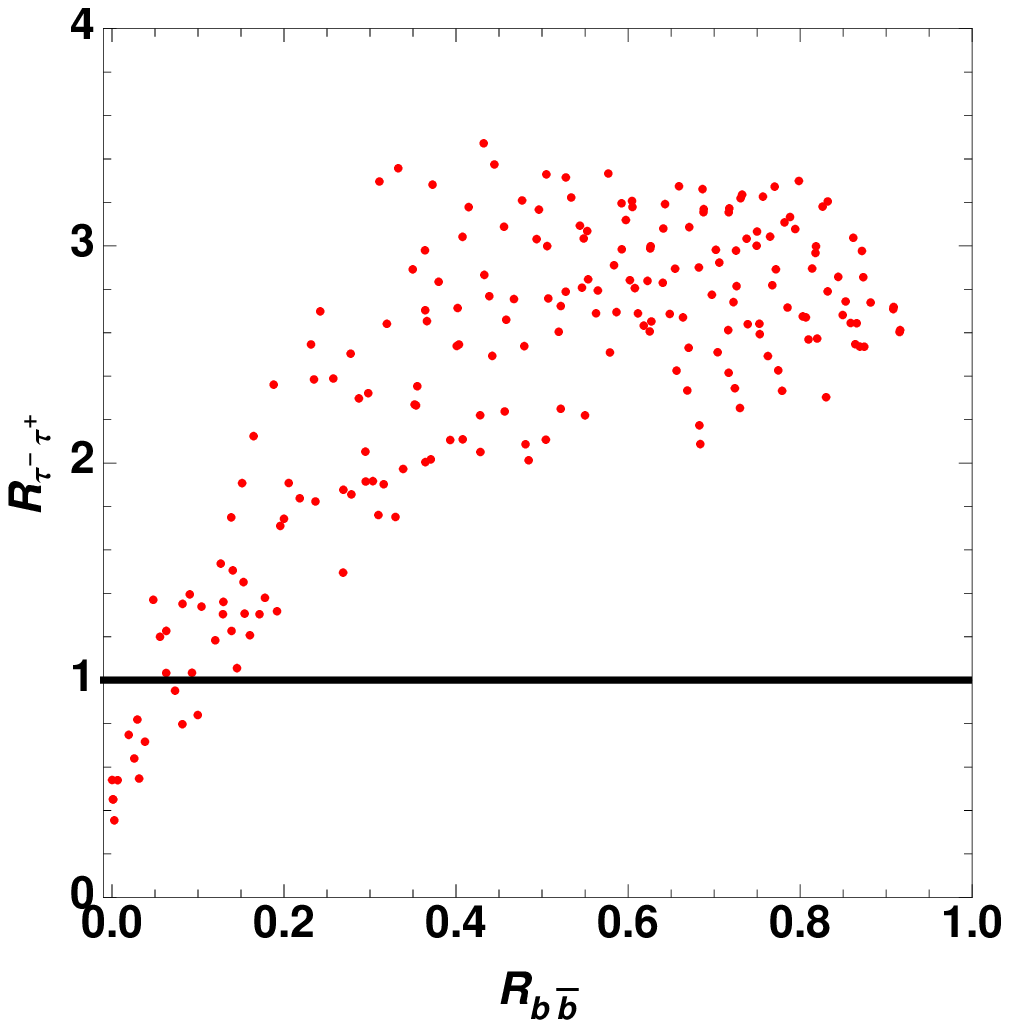}
\caption{BR$(H\to\tau^{+}\tau^{-})$ in correlation with BR$(H\to b\bar{b})$.}
\label{tautau}
\end{figure}

Discussed in \cite{hagiwara}, in the non-decoupling limit when $H\to b\bar{b}$ still dominates 
the $H$ decay, 
$H\to \tau^{+}\tau^{-}$ can be significantly enhanced.
\beq
R_{\tau\tau} \simeq r_{gg} \left(\frac{1+\Delta_{b}}{1+\Delta_{b}(1-\epsilon)}\right)^{2}
\eeq
where $\epsilon=1+\tan\alpha/\tan\beta$ with $\alpha<0$, $\Delta_{b}$ is from the radiative correction in
bottom Yukawa, $r_{gg}$ is the ratio in gluon fusion production of $H$
which is order 1 in relatively large $\tan\beta$ and $M_{\tilde{t}}>500$~GeV. With the radiative correction, $Hb\bar{b}$ coupling is
\beq
g_{Hbb}= {\cos\alpha\over \cos\beta}\left[1-\frac{\Delta_{b}}{1+\Delta_{b}}\left(1-\frac{\tan\alpha}{\tan\beta}\right)\right]~.
\eeq
Similar to the story of $\mu A_{t}$ in $b\to s$ transition, $\Delta_{b}$ also breaks Peccei-Quinn symmetry
and $R$-symmetry at the same time.  In this case, $\Delta_{b}$ contains two $R$-symmetry
breaking pieces as gluino mass $M_{\tilde{g}}$ and $A$-term contribution. Our choice of $\mu A_{t}<0$
results in cancellation between the two contribution but the enhancement to $R_{\tau\tau}$ is still significant. 
Our results also confirm the finding in \cite{hagiwara} with many points of enhanced $H\to \tau\tau$ decay.
Figure \ref{tautau} shows the correlation between BR$(H\to\tau^{+}\tau^{-})$ and BR$(H\to b\bar{b})$ in
the survival points.
On the other hand, we also find many points with $R_{\tau\tau}<1$.
One particularly interesting feature around non-decoupling limit is that $H\to h h$ decay
may open up and take significant portion of the $H$ decay.
In large parameter region, $H\to hh$ decay partial width may completely dominate the decay of $H$ once
it opens up. Discussed in \cite{hhhdecay}, the tree level $H\to hh$ decay and one loop contributions 
may have different signs and severely cancel each other\footnote{The result is based on full one loop calculation in \cite{hhhdecay} and stability of the result may require higher order calculation.}. There then exists a very fine tuned
parameter region that the $\Gamma(H\to hh)$ is at similar order as other decay and only takes about 50\%
of $H$ decay.  
If $H\to hh$ decay occurs, $h$ can further decay into $b\bar{b}$ or $\tau\tau$, the search
of $H$ then fall into the $4b$,$4\tau$ or $2b2\tau$ channels. The phenomenology of such
channels have been widely studied in the context of NMSSM with $h\to AA$ search \cite{guiyu}.
Studies of $h\to AA$ in NMSSM shows that for $M_{h}\sim 120$~GeV, it requires the 14~TeV LHC
with at least 100~fb$^{-1}$ of data to claim discovery. Therefore, we argue the $H\to hh$ decay
is not constrained by any current direct search experimental data from LHC.
In Fig.\ref{tautau}, all the points $R_{\tau\tau}<1$ bare
the same feature as BR$(H\to h h)\sim 50\%$. Among these points, predictions on $WW^{*}$ and $ZZ^{*}$ are also slightly higher than
the SM values but mostly within 1.5 which is consistent with the experimental data.
The current search of $H\to\tau\tau$ at ATLAS is still with large error bar and consistent with these large numbers
of 2 $\sigma^{\tau\tau}_{\rm SM}$. However, CMS collaboration has reported their latest data that exclude the SM $\tau\tau$ rate by 1~$\sigma$ \cite{liqiang}. If one takes this seriously, most of our final survival parameter region will be cut away and only
a few points that with significant $H\to h h$ decay can survive. In addition, the $H\to b\bar{b}$ are highly suppressed in these points and the predictions of these points agree with ATLAS central values of $R$ in all channels very well.  
In principle, the choice of $M_{A}$ can be extended to ${\cal O}(200~\text{GeV})$ in our study and the flavor bounds are less constrained for larger $M_{A}$. However, the larger $M_{A}$ region corresponds to the enhanced $R_{\tau\tau}$ region. Only smaller $M_{A}$ generates larger splitting between
$H$ and $h$ which reduces $R_{\tau\tau}$. Therefore, we only focus on the region $M_{A}\lesssim 150$~GeV.  

\begin{figure}[h]
\includegraphics[scale=1,width=7cm]{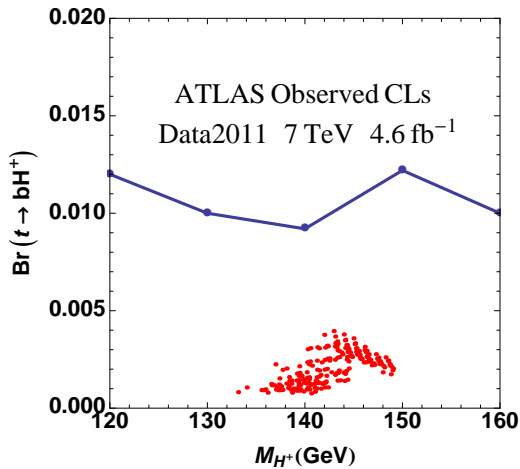}
\caption{BR$(t\to b H^{+})$ vs $M_{H^{\pm}}$ by assuming BR$(H^{+}\to \tau^{+}\nu_{\tau})=100\%$. Red dots are parameter points that pass all our selection and constraints.}
\label{tbh}
\end{figure}

Besides the direct search via $\tau\tau$, LHC has put much stronger bounds on
$t\to b H^{+}$ with $H^{+}\to \tau^{+}\nu_{\tau}$ comparing with Tevatron. The previous Tevatron upper bound of
BR$(t\to b H^{+})$ is 5\% while the latest ATLAS results become 1\%--5\%. We plot the BR$(t\to b H^{+})$ with respect
to $M_{H^{\pm}}$ by assuming BR$(H^{+}\to \tau^{+}\nu_{\tau})=100\%$ in Fig.\ref{tbh}.
It clearly shows that
all the parameter points that pass our selections are below the search of light charged Higgs boson via top decay $t\to b H^{+}$
with $H^{+}\to \tau^{+}\nu_{\tau}$.



\subsection{$\sigma_{\chi N}$}

\begin{figure}[h]
\begin{center}
\includegraphics[scale=1,width=7cm]{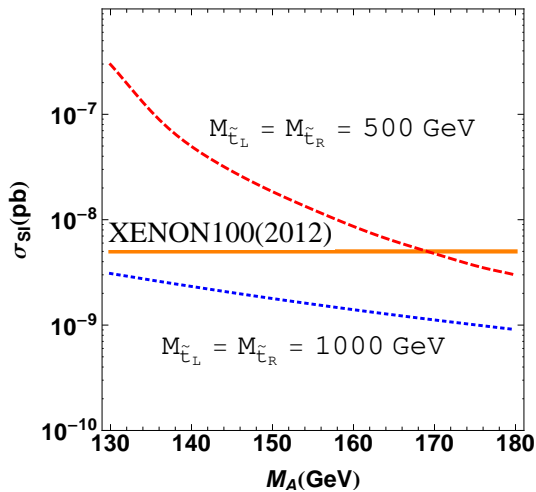}
\end{center}
\caption{Spin-independent scattering between neutralino dark matter and the nuclei computed for XENON100 by varying $M_{A}$. Red dashed line is in
the case of light stop of 500~GeV and the blue dotted line corresponds to the stop mass of 1~TeV.}
\label{dm}
\end{figure}

Finally we discuss the last constraint for non-decoupling MSSM.
Latest direct dark matter detection experiments XENON100 have reached
the level of sensitivity needed to detect neutralino dark matter over a substantial range of
supersymmetric parameter space. These experiments
attempt to detect weakly interacting (WIMP) dark matter particles through their elastic scattering with nuclei.
Neutralinos can scatter with nuclei through both scalar (spin-independent) and axial-vector
(spin-dependent) interactions. The experimental sensitivity to scalar couplings benefits from
coherent scattering, which leads to cross sections and rates proportional to the square of the
atomic mass of the target nuclei which is exactly being used for direct detection experiments.
Consequently the spin-independent interactions are far more important than the spin-dependent
in these experiments. In MSSM, the spin-independent interactions are mediated by the light Higgs bosons
with cross section proportional to
\beq
\frac{\tan^{2}\beta}{M^{4}_{A}}~.
\eeq 
Figure \ref{dm} has shown the spin-independent scattering between neutralino dark matter and the nuclei computed for XENON100 setup
by varying $M_{A}$ for pure-bino of 200~GeV to illustrate the enhancement feature due to light Higgs bosons and light top squarks. The calculation is done using {\it micrOMEGAs 2.4} \cite{micromega}. 
It clearly shows that the enhancement of such interaction in small $M_{A}$. In addition, squark can induce neutralino-gluon scattering which can further enhance the scattering cross section \cite{mihoko}. The two lines for different stop mass choices of 500~GeV and 1~TeV also indicates the enhancement of light stop in the neutralino-nuclei scattering. For most of our points with 500~GeV stop and $M_{A}\lesssim 170$~GeV,
XENON100 bounds have put stringent constraints over the scenario. On the other hand, it is not clearly whether the current dark matter completely consists
of supersymmetric neutralino. The bounds can also be easily evaded by adding new component of dark matter from non-supersymmetric origin. 

\section{Conclusions}
In this paper, we discuss the non-decoupling MSSM scenario where a light Higgs boson can evade the direct search experiments at LEP or Tevatron and
the 125~GeV Higgs-like boson is explained as the heavy Higgs boson in the spectrum. 
The light Higgs boson may evade the direct search experiments at LEPII or Tevatron while the 125~GeV Higgs-like boson is identified as 
the heavy Higgs boson in the spectrum. Two direct consequences of the scenario are the flavor violation induced by the light charged scalar and the spin-independent scattering between neutralino and nuclei in dark matter direct detection experiments. With combined flavor constraints $B\to X_{s}\gamma$ and $B_{s}\to \mu^{+}\mu^{-}$ and direct constraints on Higgs properties, we find best fit scenarios with light stop of $\cal O$(500~GeV), negative $A_{t}$ around -750~GeV and large $\mu$-term of 2-3~TeV. However, large parameter region in the survival space under all bounds may be further constrained by $H\to \tau\tau$ if no excess of $\tau\tau$ is confirmed at LHC. We only identify a small parameter region with significant $H\to hh$ decay that is consistent with all bounds and reduced $\tau\tau$ decay. In addition, if current dark matter mostly consists of neutralino, direct detection experiments like XENON100 also puts stringent bound over this scenario with light Higgs bosons. The light stops which are required by flavor constraints can further enhance the scattering cross section.

\section*{Note Added}
When completing our work, 1211.1955[hep-ph] \cite{Bechtle:2012jw} has appeared. The paper also studied similar region of non-decoupling
MSSM and the results are in agreement with ours. We also include study on its enhancement of spin-independent neutralino-nuclei
scattering. In addition, we find new parameter region which corresponds to reduce $R_{\tau\tau}$ due to $H\to hh$ decay.

\section*{Acknowledgement}
ML is supported by the National Science Foundation of China (11135006) and National Basic Research Program of China (2010CB833000). KW is supported in part, by the Zhejiang University Fundamental Research Funds for the Central Universities (2011QNA3017) and the National Science Foundation of China (11245002,11275168). GZ is supported by the National Science Foundation of China (11075139).

\end{document}